\documentclass[a4paper,twocolumn,nofootinbib,superscriptaddress]{revtex4}%
\usepackage{hyperref}
\usepackage{amsmath}
\usepackage{amsfonts}
\usepackage{amssymb}
\usepackage{graphicx}
\usepackage{color}%
\providecommand{\U}[1]{\protect\rule{.1in}{.1in}}

\begin{document}
\title{New bosonic hard string scattering amplitudes and extended Gross conjecture in
superstring theory}
\author{Sheng-Hong Lai}
\email{xgcj944137@gmail.com}
\affiliation{Department of Electrophysics, National Chiao-Tung University, Hsinchu, Taiwan, R.O.C.}
\author{Jen-Chi Lee}
\email{jcclee@cc.nctu.edu.tw}
\affiliation{Department of Electrophysics, National Chiao-Tung University, Hsinchu, Taiwan, R.O.C.}
\author{Yi Yang}
\email{yiyang@mail.nctu.edu.tw}
\affiliation{Department of Electrophysics, National Chiao-Tung University, Hsinchu, Taiwan, R.O.C.}

\begin{abstract}
We discover new hard string scattering amplitudes (HSSA) with polarizations
orthogonal to the scattering plane in the NS sector of $10D$ open superstring
theory. The corresponding HSSA in the bosonic string theory are of subleading
order in energy. In addition, all HSSA are found to be proportional to each
other and the ratios among these HSSA are calculated to justify Gross
conjecture in 1988 on high energy symmetry of string theory. Moreover,
together with these new HSSA, the total number of HSSA calculated in the first
massive level of NS sector are found to agree with those of hard polarized
fermion string scattering amplitudes (PFSSA) in the R sector calculated recently.

\end{abstract}
\maketitle

%

\setcounter{equation}{0}
\renewcommand{\theequation}{\arabic{section}.\arabic{equation}}%

\section{Introduction}

In contrast to high energy behaviour of quantum field theories, e.g. the
well-known asymptotic freedom of QCD discovered in early 70s \cite{PGW}, many
important issues of high energy behaviour of string theory \cite{GM} remained
to be studied. Historically, it was first conjectured in 1988 by Gross
\cite{Gross} in his pioneer works on this subject that there exist infinite
linear relations \cite{GrossManes} among hard string scattering amplitudes
(HSSA) of different string states in the hard scattering limit. Moreover, the
conjecture states the linear relations are so powerful that all the HSSA are
equal, up to a numerical (angle-independent) constant. This conjecture was
later corrected and proved by Taiwan group using the decoupling of zero norm
states \cite{ZNS1,2D} in \cite{ChanLee1,ChanLee2,CHL,PRL, CHLTY,susy}. For
more details, see the recent review \cite{review, over}.

In addition to the bosonic string theory, there has been attempts to extend
Gross conjecture (GC) to the NS sector of superstring theory
\cite{susy,RRsusy}. Recently, the present authors calculated a class of
polarized fermion string scattering amplitudes (PFSSA) \cite{PFSSA} at
arbitrary mass levels which involve the \textit{leading Regge trajectory}
fermion string states of the R sector, and successfully extended GC to the
fermionic sector. Since it is a nontrivial task to construct the general
massive fermion string vertex operators \cite{vertex,RRR,m1,m2,m3,m4}, the
proof of extended GC for the case of superstring theory remains an open question.

One important issue of extended GC for superstring is to compare the HSSA of
NS and R sectors at each fixed mass level of the theory. Since so far the
complete hard PFSSA is available only for the first massive level (no
non-\textit{leading Regge trajectory} fermion string states) in the R sector
\cite{PFSSA}, as a first step, in this paper we will calculate the complete
HSSA for the first massive level of the NS sector. On the other hand, we will
also calculate and count the number of HSSA for both NS and R sectors and
check whether they match to each other or not.

Since Gross conjecture was shown to be valid for both GSO even and odd states
in the NS sector \cite{susy}, we are going to ignore the GSO projection in
this paper. To be more specific, the two HSSA we want to compare are%
\begin{equation}
\left\langle V^{(-1)}Tensor^{(-1)}\phi^{(0)}\phi^{(0)}\right\rangle \label{T}%
\end{equation}
for the NS sector and%
\begin{equation}
\left\langle \psi^{(-\frac{1}{2})}\chi^{(-\frac{1}{2})}\phi^{(-1)}\phi
^{(0)}\right\rangle \label{S}%
\end{equation}
for the R sector. In Eq.(\ref{T}) the first vertex $V^{(-1)}$ is chosen to be
a massless vector, the second vertex $Tensor^{(-1)}$ is one of the two massive
tensor states and the third and the fourth vertices are tachyons $\phi^{(0)}$.
The total ghost charges sum up to $-2$. In Eq.(\ref{S}) the first vertex
$\psi^{(-\frac{1}{2})}$ is chosen to be a massless spinor, the second vertex
$\chi^{(-\frac{1}{2})}$ is the $10D$ massive Majorana spinor and the third and
the fourth vertices are tachyons $\phi^{(-1)}$ and $\phi^{(0)}$ in $(-1)$ and
$(0)$ ghost pictures respectively. Again, the total ghost charges sum up to
$-2$.

The two first vertices $\{V^{(-1)}$ $,\psi^{(-\frac{1}{2})}\}$ in Eq.(\ref{T})
and Eq.(\ref{S}) form a massless $10D$ SUSY multiplet. Similarly,
$\{Tensor^{(-1)},\chi^{(-\frac{1}{2})}\}$ form a massive SUSY multiplet
\cite{RRR} in the spectrum. The number of HSSA in Eq.(\ref{S}) was calculated
to be $16$ recently \cite{PFSSA}. On the other hand, the HSSA of Eq.(\ref{T})
with polarizations on the scattering plane was calculated in \cite{susy} and
was listed in Eq.(\ref{R}). We will use slightly different approach to obtain
the result in this paper. Since the number of bosonic HSSA calculated in
\cite{susy} was less than $16$, we expect the existence of HSSA with
polarizations orthogonal to the scattering plane. These bosonic HSSA were
mainly due to the "worldsheet fermion exchange" and were first noted in
\cite{susy}. Motivated by the recent result of PFSSA \cite{PFSSA}, we will
calculate in this paper the complete bosonic HSSA including all HSSA with
polarizations orthogonal to the scattering plane.

Recently there has been a new perspective to demonstrate the Gross conjecture
regarding the high energy symmetry of string theory, see the recent review
paper in \cite{LSSA}. The authors in \cite{LSSA} constructed the exact string
scattering amplitudes (SSA) of three tachyons and one arbitrary string state,
or the Lauricella SSA (LSSA), in the $26D$ open bosonic string theory. These
LSSA form an infinite dimensional representation of the $SL(K+3,%
\mathbb{C}
)$ group. Moreover, they show that the $SL(K+3,%
\mathbb{C}
)$ group can be used to solve all the LSSA through infinite recurrence
relations and express them in terms of $1$ amplitude. This seems to agree with
the result obtained in \cite{Mizera} that there are $((n-3)!)^{2}=1$
"linearly-independent integrals" for the case of $n=4$. Moreover, as an
application in the hard scattering limit, the LSSA can be used to rederive
infinite linear relations (with constant coefficients independent of the
scattering angle), which directly prove Gross conjecture.

In the next section, we will first calculate all physical states which include
zero norm states (ZNS) of the first mass level in the NS sector of old
covariant quantization string spectrum. In particular, we will pay attention
to ZNS and the enlarged gauge symmetries induced by them. Surprisingly, in
contrast to the open bosonic string theory \cite{ZNS1}, there is $\mathit{no}$
"inter-particle gauge transformation" for the two positive-norm states at this
mass level. Since at each fixed mass level, it is believed that string states
of both NS sector and R sector form a large gauge multiplet, this result seems
to tell us that for the case of superstring, one cannot ignore the effect of R
sector when considering high energy symmetry of superstring theory.%

\setcounter{equation}{0}
\renewcommand{\theequation}{\arabic{section}.\arabic{equation}}%

\section{ZNS and enlarged gauge transformations in the NS sector}

In this section, we will first calculate all physical states for the first
massive level in the NS sector (0 ghost picture). This will include all ZNS
and the corresponding enlarged gauge transformations induced by them.
Surprisingly, in contrast to the open bosonic string theory \cite{ZNS1}, there
is $\mathit{no}$ inter-particle enlarged gauge transformation for the
positive-norm anti-symmetric spin three state and symmetric spin two state at
this mass level. The calculation is lengthy but elementary and we will mainly
list the results.

In the old covariant first quantized (OCFQ) spectrum of 10D open superstring
theory, the solutions of physical state conditions include positive-norm
states and two types of ZNS. In the NS sector, the ZNS are \cite{GSW}%
\begin{align}
\text{Type I}  &  :G_{\frac{-1}{2}}\left\vert \chi\right\rangle ,\label{0}\\
\text{where}  &  \text{ \ }G_{\frac{1}{2}}\left\vert \chi\right\rangle
=G_{\frac{3}{2}}\left\vert \chi\right\rangle =L_{0}\left\vert \chi
\right\rangle =0;\\
& \nonumber\\
\text{Type II}  &  :\left(  G_{\frac{-3}{2}}+2G_{\frac{-1}{2}}L_{-1}\right)
\left\vert \chi^{\prime}\right\rangle ,\\
\text{where}  &  \text{ \ \ }G_{\frac{1}{2}}\left\vert \chi^{\prime
}\right\rangle =G_{\frac{3}{2}}\left\vert \chi^{\prime}\right\rangle =\left(
L_{0}+1\right)  \left\vert \chi^{\prime}\right\rangle =0.
\end{align}
In this paper, we will use the notation of reference \cite{GSW}. Note that
while type I ZNS are zero-norm at any spacetime dimension, type II ZNS are
zero-norm only at $D=10$.

The most general positive-norm states for the first massive level in the NS
sector can be calculated to be%
\begin{equation}
\left(  \epsilon_{\mu\nu\rho}b_{\frac{-1}{2}}^{\mu}b_{\frac{-1}{2}}^{\nu
}b_{\frac{-1}{2}}^{\rho}+\epsilon_{\mu\nu}b_{\frac{-1}{2}}^{\mu}\alpha
_{-1}^{\nu}+\epsilon_{\mu}b_{\frac{-3}{2}}^{\mu}\right)  \left\vert
0,k\right\rangle \label{1}%
\end{equation}
with the on-shell conditions%

\begin{equation}
3\epsilon_{\left[  \mu\nu\rho\right]  }k^{\rho}-\epsilon_{\left[  \mu
\nu\right]  }=0\text{, }\left(  \epsilon_{\left(  \mu\nu\right)  }k^{\mu
}+\epsilon_{\nu}\right)  =0.
\end{equation}
The above on-shell conditions imply that we can divide Eq.(\ref{1}) into two
states, the anti-symmetric spin three state which is described by the doublet
$\{\epsilon_{\mu\nu\rho},\epsilon_{\left[  \mu\nu\right]  }\}$%
\begin{align}
&  \left(  \epsilon_{\mu\nu\rho}b_{\frac{-1}{2}}^{\mu}b_{\frac{-1}{2}}^{\nu
}b_{\frac{-1}{2}}^{\rho}+\epsilon_{\lbrack\mu\nu]}b_{\frac{-1}{2}}^{\mu}%
\alpha_{-1}^{\nu}\right)  \left\vert 0,k\right\rangle ,\label{4}\\
&  \text{ \ with }3\epsilon_{\left[  \mu\nu\rho\right]  }k^{\rho}%
-\epsilon_{\left[  \mu\nu\right]  }=0;
\end{align}
and the symmetric spin two state which is described by the doublet
$\{\epsilon_{(\mu\nu)},\epsilon_{\mu}\}$%
\begin{align}
&  \left(  \epsilon_{(\mu\nu)}b_{\frac{-1}{2}}^{\mu}\alpha_{-1}^{\nu}%
+\epsilon_{\mu}b_{\frac{-3}{2}}^{\mu}\right)  \left\vert 0,k\right\rangle
,\label{6}\\
&  \text{ \ with }\epsilon_{\left(  \mu\nu\right)  }k^{\mu}+\epsilon_{\nu}=0.
\end{align}
\qquad

We are now ready to calculate the ZNS. The most general form of type I ZNS
turns out to be%
\begin{align}
\left\vert ZNS\right\rangle  &  =\left[  \left(  k_{[\mu}\theta_{\nu]}%
+k_{(\mu}\theta_{\nu)}-2\theta_{\left[  \mu\nu\right]  }\right)  b_{\frac
{-1}{2}}^{\mu}\alpha_{-1}^{\nu}\right. \nonumber\\
&  +\left.  k_{[\rho}\theta_{\mu\nu]}b_{\frac{-1}{2}}^{\rho}b_{\frac{-1}{2}%
}^{\mu}b_{\frac{-1}{2}}^{\nu}+\theta_{\mu}b_{\frac{-3}{2}}^{\mu}\right]
\left\vert 0,k\right\rangle \label{2}%
\end{align}
with the on-shell conditions%
\begin{equation}
2k^{\mu}\theta_{\left[  \mu\nu\right]  }+\theta_{\nu}=0,(\text{automatically
}k\cdot\theta=0). \label{7}%
\end{equation}
For the rest of this paper, we define $e^{P}=\frac{1}{M_{2}}(E_{2}%
,\mathrm{k}_{2},0)=\frac{k_{2}}{M_{2}}$ the momentum polarization,
$e^{L}=\frac{1}{M_{2}}(\mathrm{k}_{2},E_{2},0)$ the longitudinal polarization
and $e^{T}=(0,0,1)$ the transverse polarization on the scattering plane. In
addition, we define $e^{T_{I}}=\{e^{T},e^{T_{3}},e^{T_{4}},....,e^{T_{9}%
})=\{e^{T},e^{T_{i}}\}$ to represent all transverse directions including
directions $e^{T_{i}}$ orthogonal to the scattering plane.

One obvious solution of Eq.(\ref{7}) is to choose%
\begin{equation}
\theta_{\nu}^{\left(  1\right)  }=0\text{ and }\theta_{\left[  \mu\nu\right]
}^{\left(  1\right)  }=\frac{1}{2}\left(  e_{\mu}^{I}e_{\nu}^{J}-e_{\nu}%
^{I}e_{\mu}^{J}\right)  \text{, }I,J\neq P\text{,}%
\end{equation}
and one obtains the anti-symmetric spin two ZNS%
\begin{equation}
\left\vert ZNS\right\rangle =\left(  k_{[\rho}\theta_{\mu\nu]}b_{\frac{-1}{2}%
}^{\rho}b_{\frac{-1}{2}}^{\mu}b_{\frac{-1}{2}}^{\nu}-2\theta_{\left[  \mu
\nu\right]  }b_{\frac{-1}{2}}^{\mu}\alpha_{-1}^{\nu}\right)  \left\vert
0,k\right\rangle , \label{3}%
\end{equation}
with $k^{\mu}\theta_{\left[  \mu\nu\right]  }=0$. It can be shown that the ZNS
in Eq.(\ref{3}) generates an enlarged on-shell gauge transformation%
\begin{equation}
\epsilon_{\left[  \mu\nu\rho\right]  }^{\prime}=\left(  \epsilon_{\left[
\mu\nu\rho\right]  }+k_{[\mu}\theta_{\nu\rho]}\right)  ,\epsilon_{\left[
\mu\nu\right]  }^{\prime}=\epsilon_{\left[  \mu\nu\right]  }-\theta_{\left[
\mu\nu\right]  }%
\end{equation}
for the anti-symmetric spin three state in Eq.(\ref{4}).

The second solution of Eq.(\ref{7}) is to choose%
\begin{align}
\theta_{\nu}^{\left(  2\right)  }  &  =-2k^{\mu}\theta_{\left[  \mu\nu\right]
}=\frac{-1}{M}k^{\mu}\left(  e_{\mu}^{P}e_{\nu}^{K}-e_{\nu}^{P}e_{\mu}%
^{K}\right)  =e_{\nu}^{K},\\
\theta_{\left[  \mu\nu\right]  }^{\left(  2\right)  }  &  =\frac{1}{2M}\left(
e_{\mu}^{P}e_{\nu}^{K}-e_{\nu}^{P}e_{\mu}^{K}\right)  \text{, }K\neq P.
\end{align}
For this choice, Eq.(\ref{2}) got simplified and one obtains a vector ZNS%
\begin{equation}
\left\vert ZNS\right\rangle =\left(  k_{(\mu}\theta_{\nu)}b_{\frac{-1}{2}%
}^{\mu}\alpha_{-1}^{\nu}+\theta_{\mu}b_{\frac{-3}{2}}^{\mu}\right)  \left\vert
0,k\right\rangle , \label{8}%
\end{equation}
with $k\cdot\theta=0$. It can be shown that the ZNS in Eq.(\ref{8}) generates
an enlarged on-shell gauge transformation%
\begin{equation}
\epsilon_{\left(  \mu\nu\right)  }^{\prime}=\epsilon_{\left(  \mu\nu\right)
}+k_{(\mu}\theta_{\nu)},\epsilon_{\mu}^{\prime}=\left[  \epsilon_{\mu}%
+\theta_{\mu}\right]
\end{equation}
for the symmetric spin two state in Eq.(\ref{6}). To obtain the vector ZNS
state in Eq.(\ref{8}), one can alternatively propose%
\begin{equation}
\left\vert \chi\right\rangle =\left[  \frac{1}{2M^{2}}\left(  k_{\mu}%
\theta_{\nu}-k_{\nu}\theta_{\mu}\right)  b_{\frac{-1}{2}}^{\mu}b_{\frac{-1}%
{2}}^{\nu}+\theta_{\mu}\alpha_{-1}^{\mu}\right]  \left\vert 0,k\right\rangle ,
\end{equation}
with $k\cdot\theta=0$ and put it in Eq.(\ref{0}).

The only type II ZNS is a scalar and can be calculated to be%
\begin{equation}
\left\vert ZNS\right\rangle =\left[  3k_{\mu}b_{\frac{-3}{2}}^{\mu}+\left(
\eta_{\mu\nu}+2k_{\mu}k_{\nu}\right)  b_{\frac{-1}{2}}^{\mu}\alpha_{-1}^{\nu
}\right]  \left\vert 0,k\right\rangle . \label{5}%
\end{equation}
It can be shown that the ZNS in Eq.(\ref{5}) generates an enlarged on-shell
gauge transformation%
\begin{equation}
\epsilon_{\left(  \mu\nu\right)  }^{\prime}=\epsilon_{\left(  \mu\nu\right)
}+\eta_{\mu\nu}+2k_{\mu}k_{\nu},\epsilon_{\mu}^{\prime}=\epsilon_{\mu}%
+3k_{\mu}%
\end{equation}
for the symmetric spin two state in Eq.(\ref{6}).

We note that the number of ZNS in Eq.(\ref{3}), Eq.(\ref{8}) and Eq.(\ref{5})
can be counted to be $36+9+1=46$. On the other hand, the number of "spurious
states" in the $-1$ ghost picture were calculated \cite{RRR} to be
$36+9+1=46,$ which is consistent with the results calculated in the $0$ ghost
picture in this paper.

Surprisingly, in contrast to the open bosonic string theory \cite{ZNS1}, there
is $\mathit{no}$ "inter-particle gauge transformation" for the positive-norm
anti-symmetric spin three state and symmetric spin two state at this mass level.

Here we briefly review the results of $26D$ open bosonic string theory at mass
level $M^{2}=4$ \cite{ZNS1}. At this mass level, there are two positive-norm
states, the symmetric spin three state and the anti-symmetric spin two state.
In addition there are three type I ZNS, the symmetric spin two, a vector and a
scalar, and one type II vector ZNS. One can do a linear combination of the two
degenerate vector ZNS to obtain the so-called $D_{2}$ vector ZNS \cite{ZNS1}%
\begin{align}
|D_{2}\rangle &  =\left[  \left(  \frac{1}{2}k_{\mu}k_{\nu}\theta_{\lambda
}^{2}+2\eta_{\mu\nu}\theta_{\lambda}^{2}\right)  \alpha_{-1}^{\mu}\alpha
_{-1}^{\nu}\alpha_{-1}^{\lambda}\right. \nonumber\\
&  +\left.  9k_{\mu}\theta_{\nu}^{2}\alpha_{-2}^{[\mu}\alpha_{-1}^{\nu
]}-6\theta_{\mu}^{2}\alpha_{-3}^{\mu}\right]  \left\vert 0,k\right\rangle ,
\label{02}%
\end{align}
with $k\cdot\theta^{2}=0$. The above $D_{2}$ ZNS generates an "inter-particle
gauge transformation" for the positive-norm symmetric spin three state and the
anti-symmetric spin two state. Thus these two states form a large gauge
multiplet. The other three ZNS generate gauge transformations for the
symmetric spin three state. An important consequence of this inter-particle
symmetry is the linear relation which relates hard string scattering
amplitudes of the two positive-norm states. Indeed, all HSSA are proportional
to each others with the ratios \cite{ChanLee1,ChanLee2}%
\begin{align}
&  \mathcal{T}_{TTT}:\mathcal{T}_{LLT}:\mathcal{T}_{(LT)}:\mathcal{T}%
_{[LT]}\nonumber\\
=  &  \alpha_{-1}^{T}\alpha_{-1}^{T}\alpha_{-1}^{T}:\alpha_{-1}^{L}\alpha
_{-1}^{L}\alpha_{-1}^{T}:\alpha_{-1}^{(L}\alpha_{-2}^{T)}:\alpha_{-1}%
^{[L}\alpha_{-2}^{T]}\nonumber\\
=  &  8:1:-1:-1. \label{bosonic}%
\end{align}
Although there is no "inter-particle gauge transformation" at mass level
$M^{2}=2$ in the NS sector of the spectrum of the superstring, in the next
section we will see that the $16$ HSSA originated from the two positive-norm
states are still related and are indeed again proportional to each others.
Presumably, this is due to the spacetime SUSY and the spacetime massive
fermion string scattering amplitudes of the fermionic sector of the theory
\cite{PFSSA}.

\section{Hard string scattering amplitudes in the NS sector}

In this section we will calculate all HSSA in Eq.(\ref{T}) including some
$\mathit{new}$ ones which were not considered in \cite{susy}. We will
demonstrate the linear relations among the HSSA between the two positive-norm
states, and thus extend and justify the original Gross conjecture to the case
of superstring. Similar inter-particle symmetry was discovered
\cite{ChanLee1,ChanLee2} in the 26D bosonic string theory.

For the case of the bosonic string, string scattering amplitudes with
polarizations orthogonal to the scattering plane are of subleading order in
energy in the hard scattering limit. However, it was noted that \cite{susy}
for the HSSA of the NS sector of superstring, there existed leading order HSSA
with polarizations orthogonal to the scattering plane. This was due to the
"worldsheet fermion exchange" in the correlation functions and was argued to
be related to the massive spacetime fermion HSSA of the theory.

In the subsection A of this section, we will first calculate ratios among HSSA
with polarizations on the scattering plane by using the hard Ward identities
(or decoupling of ZNS in the hard scattering limit), and then by using
explicit HSSA calculations. In subsections B and C, we will calculate two
types of new HSSA with polarizations orthogonal to the scattering plane.

\subsection{Hard SSA with polarizations on the scattering plane}

We first do the hard Ward identity (HWI) calculation for polarizations on the
scattering plane. In the hard scattering limit, one identifies $e^{L}=e^{P}$
\cite{ChanLee1,ChanLee2}. In the following calculation, each scattering
amplitude has been assigned a relative energy power. For each longitudinal $L$
component, the energy order is $E^{2}$ while for each transverse $T$
component, the energy order is $E.$ This is due to the definitions of $e^{L}$
and $e^{T}$ in the paragraph below Eq.(\ref{7}) where $e^{L}$ got one energy
power more than that of $e^{T}.$

In the HWI calculation, we put the ZNS in the second vertex and suppress the
indices for the other vertices. The ZNS in Eq.(\ref{3}) gives the HWI%
\begin{equation}
\left[  \overset{\left(  3\rightarrow1\right)  }{\overbrace{\alpha_{-1}%
^{L}b_{\frac{-1}{2}}^{T}-\alpha_{-1}^{T}b_{\frac{-1}{2}}^{L}}}\right]
\left\vert 0,k\right\rangle =0. \label{a}%
\end{equation}
For the choice of $\theta^{\left(  1\right)  }=e^{L}$, the ZNS in Eq.(\ref{8})
gives the HWI%
\begin{equation}
\left[  \overset{\left(  2\right)  }{\overbrace{b_{\frac{-3}{2}}^{L}}}%
+\frac{1}{M}\overset{\left(  4\rightarrow2\right)  }{\overbrace{\alpha
_{-1}^{L}b_{\frac{-1}{2}}^{L}}}\right]  \left\vert 0,k\right\rangle .
\label{b}%
\end{equation}
For the choice of $\theta^{\left(  1\right)  }=e^{T}$, the ZNS in Eq.(\ref{8})
gives the HWI%
\begin{equation}
\left[  \overset{1}{\overbrace{b_{\frac{-3}{2}}^{T}}}+\frac{1}{2M}%
\overset{\left(  3\rightarrow1\right)  }{\overbrace{\left(  \alpha_{-1}%
^{L}b_{\frac{-1}{2}}^{T}+\alpha_{-1}^{T}b_{\frac{-1}{2}}^{L}\right)  }%
}\right]  \left\vert 0,k\right\rangle =0. \label{c}%
\end{equation}
The ZNS in Eq.(\ref{5}) gives the HWI%
\begin{equation}
\left(  \frac{3}{M}b_{\frac{-3}{2}}^{L}+\eta_{\mu\nu}\alpha_{-1}^{\mu}%
b_{\frac{-1}{2}}^{\nu}+\frac{2}{M^{2}}\alpha_{-1}^{L}b_{\frac{-1}{2}}%
^{L}\right)  \left\vert 0,k\right\rangle =0
\end{equation}
where%
\begin{equation}
\eta_{\mu\nu}=-e_{\mu}^{P}e_{\nu}^{P}+e_{\mu}^{L}e_{\nu}^{L}+e_{\mu}^{T}%
e_{\nu}^{T}+\underset{\alpha_{-1}^{T_{i}}b_{\frac{1}{2}}^{T_{i}}%
\rightarrow\text{not contribute}}{\underbrace{\overset{7}{\underset{i=1}{\sum
}}e_{\mu}^{T_{i}}e_{\nu}^{T_{i}}}.}%
\end{equation}
In sum, ZNS in Eq.(\ref{5}) gives the HWI%
\begin{equation}
\left[  3M\overset{\left(  2\right)  }{\overbrace{b_{\frac{-1}{2}}^{L}}}%
+M^{2}\overset{\left(  2\right)  }{\overbrace{\alpha_{-1}^{T}b_{\frac{-1}{2}%
}^{T}}}+2\overset{\left(  4\rightarrow2\right)  }{\overbrace{\alpha_{-1}%
^{L}b_{\frac{-1}{2}}^{L}}}\right]  \left\vert 0,k\right\rangle =0. \label{d}%
\end{equation}
The naive energy order of $2\alpha_{-1}^{L}b_{\frac{-1}{2}}^{L}$ term in
Eq.(\ref{d}) is $E^{4}$, the HWI forces its energy order to drop to $E^{2}$.
Similar calculations were performed in Eq.(\ref{a}),Eq.(\ref{b}) and
Eq.(\ref{c}). We note that the energy order $E^{3}$ term in Eq.(\ref{c}) gives%
\begin{equation}
\alpha_{-1}^{L}b_{\frac{-1}{2}}^{T}+\alpha_{-1}^{T}b_{\frac{-1}{2}}^{L}=0,
\end{equation}
which together with Eq.(\ref{a}) force the following two $E^{3}$ order terms
vanish%
\begin{equation}
\overset{\left(  3\right)  }{\overbrace{\alpha_{-1}^{L}b_{\frac{-1}{2}}^{T}}%
}=0=\overset{\left(  3\right)  }{\overbrace{\alpha_{-1}^{T}b_{\frac{-1}{2}%
}^{L}}}.
\end{equation}
Thus the real leading order terms are those of energy order $E^{2}$ in
Eq.(\ref{b}) and Eq.(\ref{d})%
\begin{align}
\alpha_{-1}^{L}b_{\frac{-1}{2}}^{L}+Mb_{\frac{-3}{2}}^{L}  &  =0,\\
3Mb_{\frac{-1}{2}}^{L}+M^{2}\alpha_{-1}^{T}b_{\frac{-1}{2}}^{T}+2\alpha
_{-1}^{L}b_{\frac{-1}{2}}^{L}  &  =0,
\end{align}
which can be solved to give the ratios%
\begin{equation}
b_{\frac{-3}{2}}^{L}:\alpha_{-1}^{L}b_{\frac{-1}{2}}^{L}:\alpha_{-1}%
^{T}b_{\frac{-1}{2}}^{T}=\frac{-1}{M}:\frac{1}{M^{2}}:1. \label{R}%
\end{equation}
The result in Eq.(\ref{R}) is consistent with the calculation obtained in
\cite{susy} by a slightly different approach.

The ratios in Eq.(\ref{R}) can be rederived by calculating a set of sample
HSSA. We first calculate%
\begin{align}
&  \left\langle V^{(-1)}Tensor^{(-1)}\phi^{(0)}\phi^{(0)}\right\rangle =%
{\displaystyle\int}
dz_{1}dz_{2}dz_{3}dz_{4}\frac{\left\vert z_{13}z_{14}z_{34}\right\vert
}{dz_{1}dz_{3}dz_{4}}\nonumber\\
&  \cdot\left\langle
\begin{array}
[c]{c}%
\left(  \epsilon_{1}^{\mu_{1}}\psi_{1\mu_{1}}e^{-\phi_{1}}e^{ik_{1}X_{1}%
}\right)  \left(  \epsilon_{2}^{\mu_{2}}\psi_{2\mu_{2}}e^{-\phi_{2}}%
i\epsilon_{\mu}\partial X_{2}^{\mu}e^{ik_{2}X_{2}}\right) \\
\times\left(  k_{3}^{\mu_{3}}\psi_{3\mu_{3}}e^{ik_{3}X_{3}}\right)  \left(
k_{4}^{\mu_{4}}\psi_{4\mu_{4}}e^{ik_{4}X_{4}}\right)
\end{array}
\right\rangle
\end{align}
where the first vertex $V^{(-1)}$ is chosen to be a vector, the second vertex
is the $\alpha_{-1}^{T}b_{-\frac{1}{2}}^{T}\left\vert 0;k\right\rangle $
tensor state and the third and the fourth vertices are tachyons $\phi^{(0)}$.
The total ghost charges sum up to $-2$. We first put $\epsilon=e^{T}$ to get
\begin{widetext}
\begin{align}
\left\langle V^{(-1)}Tensor^{(-1)}\phi^{(0)}\phi^{(0)}\right\rangle  &
=\epsilon_{1}^{\mu_{1}}\epsilon_{2}^{\mu_{2}}k_{3}^{\mu_{3}}k_{4}^{\mu_{4}}%
{\displaystyle\int}
dz_{2}\left[  \frac{\eta_{\mu_{1}\mu_{2}}}{z_{12}}\frac{\eta_{\mu_{3}\mu_{4}}%
}{z_{34}}-\frac{\eta_{\mu_{1}\mu_{3}}}{z_{13}}\frac{\eta_{\mu_{2}\mu_{4}}%
}{z_{24}}+\frac{\eta_{\mu_{1}\mu_{4}}}{z_{14}}\frac{\eta_{\mu_{2}\mu_{3}}%
}{z_{23}}\right]  \nonumber\\
&  \times\left\vert z_{12}\right\vert ^{k_{1}\cdot k_{2}-1}\left\vert
z_{13}\right\vert ^{k_{1}\cdot k_{3}+1}\left\vert z_{14}\right\vert
^{k_{1}\cdot k_{4}+1}\left\vert z_{23}\right\vert ^{k_{2}\cdot k_{3}%
}\left\vert z_{24}\right\vert ^{k_{2}\cdot k_{4}}\left\vert z_{34}\right\vert
^{k_{3}\cdot k_{4}+1}\left(  \frac{k_{3}^{T}}{z_{23}}+\frac{k_{4}^{T}}{z_{24}%
}\right)  .
\end{align}
\end{widetext}The next step is to fix the gauge $z_{1}=0,z_{3}=1,z_{4}%
\rightarrow\infty$ $(0<z_{2}<1)$ and use $k_{1}\cdot k_{2}=-\frac{s}%
{2}+1,k_{2}\cdot k_{3}=-\frac{t}{2}+\frac{1}{2}$ to perform the integration to
obtain
\begin{align}
&  \left\langle V^{(-1)}Tensor^{(-1)}\phi^{(0)}\phi^{(0)}\right\rangle
=B\left(  \frac{-s}{2},\frac{-t}{2}+\frac{1}{2}\right) \nonumber\\
&  \cdot\left(  -k_{3}^{T}\right)  \left[
\begin{array}
[c]{c}%
\left(  \epsilon_{1}\cdot\epsilon_{2}\right)  \left(  k_{3}\cdot k_{4}\right)
\\
-\left(  k_{3}\cdot\epsilon_{1}\right)  \left(  k_{4}\cdot\epsilon_{2}\right)
\frac{s}{u+1}\\
+\left(  k_{3}\cdot\epsilon_{2}\right)  \left(  k_{4}\cdot\epsilon_{1}\right)
\frac{s}{t+1}%
\end{array}
\right]  . \label{same}%
\end{align}
where $B\left(  \frac{-s}{2},\frac{-t}{2}+\frac{1}{2}\right)  =\frac
{\Gamma\left(  \frac{-s}{2}\right)  \Gamma\left(  \frac{-t}{2}+\frac{1}%
{2}\right)  }{\Gamma\left(  \frac{u}{2}+\frac{1}{2}\right)  }$ is the Beta function.

Finally in the hard scattering limit we have%
\begin{align}
k_{1}  &  =\left(  E,-E,0\right)  ,k_{2}=\left(  E,E,0\right)  ,\nonumber\\
k_{3}  &  =\left(  -E,-E\cos\theta,-E\sin\theta\right)  ,\nonumber\\
k_{4}  &  =\left(  -E,E\cos\theta,E\sin\theta\right)
\end{align}
and%
\begin{align}
\epsilon_{1}  &  =\epsilon_{2}=e^{T}=\left(  0,0,1\right)  ,s=4E^{2}%
,\nonumber\\
\frac{s}{u}  &  \rightarrow\frac{-1}{\cos^{2}\frac{\theta}{2}},\frac{s}%
{t}\rightarrow\frac{-1}{\sin^{2}\frac{\theta}{2}},k_{3}\cdot k_{4}=\frac
{-s}{2}-1.
\end{align}
After a bit calculation, one ends up with%
\begin{equation}
\left\langle V^{(-1)}Tensor_{(\alpha_{-1}^{T}b_{\frac{-1}{2}}^{T})}^{(-1)}%
\phi^{(0)}\phi^{(0)}\right\rangle =-2k_{3}^{T}E^{2}B \label{h1}%
\end{equation}
where $B$ is the hard scattering limit of the Beta function.

Similar calculations can be performed if one replaces the second vertex
$\alpha_{-1}^{T}b_{\frac{-1}{2}}^{T}$ by $\alpha_{-1}^{L}b_{\frac{-1}{2}}^{L}%
$and $b_{-\frac{3}{2}}^{T}\left\vert 0;k\right\rangle $ respectively. The
results are%
\begin{align}
\left\langle V^{(-1)}Tensor_{(\alpha_{-1}^{L}b_{\frac{-1}{2}}^{L})}^{(-1)}%
\phi^{(0)}\phi^{(0)}\right\rangle  &  =\frac{-2k_{3}^{T}E^{2}B}{M^{2}%
},\label{h2}\\
\left\langle V^{(-1)}Tensor_{(b_{-\frac{3}{2}}^{T})}^{(-1)}\phi^{(0)}%
\phi^{(0)}\right\rangle  &  =\frac{2k_{3}^{T}E^{2}B}{M}. \label{h3}%
\end{align}
These results justify the ratios calculated in Eq.(\ref{R}). They also agree
with the results obtained by the saddle-point method \cite{susy}.

\subsection{New hard SSA I}

In addition to the three HSSA calculated in Eq.(\ref{h1}), Eq.(\ref{h2}) and
Eq.(\ref{h3}), it was noted that \cite{susy} there existed HSSA with
polarizations orthogonal to the scattering plane. In particular, one can
replace $b_{\frac{-1}{2}}^{T}$ operator in the vertex by $b_{\frac{-1}{2}%
}^{T_{i}}$where $T_{i},i=3,4,5,\cdots,9$ represents directions orthogonal to
the scattering plane. Note that the corresponding string scattering amplitudes
are of subleading order in energy for the case of bosonic string theory. We
first propose the following HSSA (we use the obvious brief notation)%
\begin{align}
&  \left\langle b_{\frac{-1}{2}}^{T_{i}},\alpha_{-1}^{T}b_{\frac{-1}{2}%
}^{T_{j}},\phi_{3},\phi_{4}\right\rangle =B\left(  \frac{-s}{2},\frac{-t}%
{2}+\frac{1}{2}\right) \nonumber\\
&  \cdot\left(  -k_{3}^{T}\right)  \left[
\begin{array}
[c]{c}%
\left(  e^{T_{i}}\cdot e^{T_{j}}\right)  \left(  k_{3}\cdot k_{4}\right) \\
-\left(  k_{3}\cdot e^{T_{i}}\right)  \left(  k_{4}\cdot e^{T_{j}}\right)
\frac{s}{u+1}\\
+\left(  k_{3}\cdot e^{T_{j}}\right)  \left(  k_{4}\cdot e^{T_{i}}\right)
\frac{s}{t+1}%
\end{array}
\right]  .
\end{align}
The calculation of the above amplitude is similar to Eq.(\ref{same}).

Finally in the hard scattering limit, one ends up with%
\begin{equation}
\left\langle b_{\frac{-1}{2}}^{T_{i}},\alpha_{-1}^{T}b_{\frac{-1}{2}}^{T_{j}%
},\phi_{3},\phi_{4}\right\rangle =\delta_{ij}2k_{3}^{T}E^{2}B. \label{more}%
\end{equation}
Eq.(\ref{more}) contributes $7$ more HSSA in the NS sector at this mass level.

\subsection{New hard SSA II}

So far all HSSA calculated in this section are from the positive-norm
symmetric spin two state. One expects HSSA contributed from the anti-symmetric
spin three state. In this subsection we propose the following HSSA%
\begin{equation}
\left\langle b_{\frac{-1}{2}}^{T_{K}},b_{\frac{-1}{2}}^{L}b_{\frac{-1}{2}%
}^{T_{I}}b_{\frac{-1}{2}}^{T_{J}},\phi_{3},\phi_{4}\right\rangle ,
\end{equation}
which was not considered in \cite{susy}. The calculation of this HSSA is
lengthy but straightforward. We begin with the standard calculation of string
scattering amplitude\begin{widetext}
\begin{align}
&  \left\langle V^{(-1)}Tensor_{bbb}^{(-1)}\phi^{(0)}\phi^{(0)}\right\rangle =%
{\displaystyle\int}
dz_{1}dz_{2}dz_{3}dz_{4}\frac{\left\vert z_{13}z_{14}z_{34}\right\vert
}{dz_{1}dz_{3}dz_{4}}\left\langle
\begin{array}
[c]{c}%
\left(  \epsilon_{\mu_{1}}^{T_{K}}\psi_{1}^{\mu_{1}}e^{-\phi_{1}}%
e^{ik_{1}X_{1}}\right)  \left(  \epsilon_{\mu_{2}}^{L}\epsilon_{\mu3}^{T_{I}%
}\epsilon_{\mu4}^{T_{J}}\psi_{2}^{\mu_{2}}\psi_{2}^{\mu_{3}}\psi_{2}^{\mu_{4}%
}e^{-\phi_{1}}e^{ik_{2}X_{2}}\right)  \\
\times\left(  k_{\mu_{5}}^{3}\psi_{3}^{\mu_{5}}e^{ik_{3}X_{3}}\right)  \left(
k_{\mu_{6}}^{4}\psi_{4}^{\mu_{6}}e^{ik_{4}X_{4}}\right)
\end{array}
\right\rangle \nonumber\\
&  =\epsilon_{\mu_{1}}^{T_{K}}\epsilon_{\mu_{2}}^{L}\epsilon_{\mu3}^{T_{I}%
}\epsilon_{\mu4}^{T_{J}}k_{\mu_{5}}^{3}k_{\mu_{6}}^{4}%
{\displaystyle\int}
dz_{2}\left\vert z_{13}z_{14}z_{34}\right\vert \left\langle \psi_{1}^{\mu_{1}%
}\psi_{2}^{\mu_{2}}\psi_{2}^{\mu_{3}}\psi_{2}^{\mu_{4}}\psi_{3}^{\mu_{5}}%
\psi_{4}^{\mu_{6}}\right\rangle \left\langle e^{-\phi_{1}}e^{-\phi_{2}%
}\right\rangle \left\langle e^{ik_{1}X_{1}}e^{ik_{2}X_{2}}e^{ik_{3}X_{3}%
}e^{ik_{4}X_{4}}\right\rangle . \label{two}
\end{align}
\end{widetext}The next step is to fix the gauge $z_{1}=0,z_{3}=1,z_{4}%
\rightarrow\infty$ $(0<z_{2}<1)$ to get%
\begin{align}
&  \left\langle V^{(-1)}Tensor_{bbb}^{(-1)}\phi^{(0)}\phi^{(0)}\right\rangle
\nonumber\\
= &  -\epsilon_{\mu_{1}}^{T_{K}}\epsilon_{\mu_{2}}^{L}\epsilon_{\mu3}^{T_{I}%
}\epsilon_{\mu4}^{T_{J}}k_{\mu_{5}}^{3}k_{\mu_{6}}^{4}%
{\displaystyle\int}
dz_{2}\left(  z_{2}\right)  ^{k_{1}\cdot k_{2}-2}\left(  1-z_{2}\right)
^{k_{2}\cdot k_{3}-1}\nonumber\\
&  \cdot\left(
\begin{array}
[c]{c}%
\eta^{\mu_{1}\mu_{2}}\eta^{\mu_{4}\mu_{5}}\eta^{\mu_{3}\mu_{6}}-\eta^{\mu
_{1}\mu_{2}}\eta^{\mu_{3}\mu_{5}}\eta^{\mu_{4}\mu_{6}}\\
\eta^{\mu_{1}\mu_{3}}\eta^{\mu_{2}\mu_{5}}\eta^{\mu_{4}\mu_{6}}-\eta^{\mu
_{1}\mu_{3}}\eta^{\mu_{4}\mu_{5}}\eta^{\mu_{2}\mu_{6}}\\
\eta^{\mu_{1}\mu_{4}}\eta^{\mu_{3}\mu_{5}}\eta^{\mu_{2}\mu_{6}}-\eta^{\mu
_{1}\mu_{4}}\eta^{\mu_{2}\mu_{5}}\eta^{\mu_{3}\mu_{6}}%
\end{array}
\right)  .
\end{align}
We can now use $k_{1}\cdot k_{2}=-\frac{s}{2}+1,k_{2}\cdot k_{3}=-\frac{t}%
{2}+\frac{1}{2}$ and perform the integration to get%
\begin{align}
&  \left\langle V^{(-1)}Tensor_{bbb}^{(-1)}\phi^{(0)}\phi^{(0)}\right\rangle
\nonumber\\
= &  -\epsilon_{\mu_{1}}^{T_{K}}\epsilon_{\mu_{2}}^{L}\epsilon_{\mu3}^{T_{I}%
}\epsilon_{\mu4}^{T_{J}}k_{\mu_{5}}^{3}k_{\mu_{6}}^{4}%
{\displaystyle\int}
dz_{2}\left(  z_{2}\right)  ^{-\frac{s}{2}-1}\left(  1-z_{2}\right)
^{-\frac{t}{2}-\frac{1}{2}}\nonumber\\
&  \cdot\left(
\begin{array}
[c]{c}%
\eta^{\mu_{1}\mu_{3}}\eta^{\mu_{2}\mu_{5}}\eta^{\mu_{4}\mu_{6}}-\eta^{\mu
_{1}\mu_{3}}\eta^{\mu_{4}\mu_{5}}\eta^{\mu_{2}\mu_{6}}\\
\eta^{\mu_{1}\mu_{4}}\eta^{\mu_{3}\mu_{5}}\eta^{\mu_{2}\mu_{6}}-\eta^{\mu
_{1}\mu_{4}}\eta^{\mu_{2}\mu_{5}}\eta^{\mu_{3}\mu_{6}}%
\end{array}
\right)  .\nonumber\\
= &  -\frac{\Gamma\left(  \frac{-s}{2}\right)  \Gamma\left(  \frac{-t}%
{2}+\frac{1}{2}\right)  }{\Gamma\left(  \frac{u}{2}+\frac{1}{2}\right)
}\nonumber\\
&  \cdot\left(
\begin{array}
[c]{c}%
\delta_{IK}\frac{1}{M}\left(  \frac{-t}{2}+\frac{1}{2}\right)  k_{4}^{T_{J}%
}-\delta_{IK}\frac{1}{M}\left(  \frac{-u}{2}+\frac{1}{2}\right)  k_{3}^{T_{J}%
}\\
\delta_{JK}\frac{1}{M}\left(  \frac{-u}{2}+\frac{1}{2}\right)  k_{3}^{T_{I}%
}-\delta_{JK}\frac{1}{M}\left(  \frac{-t}{2}+\frac{1}{2}\right)  k_{4}^{T_{I}}%
\end{array}
\right)  .
\end{align}
In the hard scattering limit, the HSSA reduces to%
\begin{align}
&  \left\langle b_{\frac{-1}{2}}^{T_{K}},b_{\frac{-1}{2}}^{L}b_{\frac{-1}{2}%
}^{T_{I}}b_{\frac{-1}{2}}^{T_{J}},\phi_{3},\phi_{4}\right\rangle \nonumber\\
= &  -\left[
\begin{array}
[c]{c}%
\delta_{IK}\frac{1}{M}\left(  \frac{-t}{2}\right)  k_{4}^{T_{J}}-\delta
_{IK}\frac{1}{M}\left(  \frac{-u}{2}\right)  k_{3}^{T_{J}}\\
\delta_{JK}\frac{1}{M}\left(  \frac{-u}{2}\right)  k_{3}^{T_{I}}-\delta
_{JK}\frac{1}{M}\left(  \frac{-t}{2}\right)  k_{4}^{T_{I}}%
\end{array}
\right]  B.
\end{align}
There are three cases to further reduce the result:

\textbf{Case one}: for $\epsilon^{T_{K}}=e^{T},\epsilon^{T_{I}}=e^{T}%
,\epsilon^{T_{J}}=e^{T_{j}},$%
\begin{align}
&  -\left[  \frac{1}{M}\left(  \frac{-t}{2}\right)  k_{4}^{T_{j}}-\frac{1}%
{M}\left(  \frac{-u}{2}\right)  k_{3}^{T_{j}}+0-0\right] \nonumber\\
&  =\frac{1}{M}\left(  \frac{t}{2}\right)  \times0+\frac{1}{M}\left(
\frac{-u}{2}\right)  \times0=0.
\end{align}

\textbf{Case two}: for $\epsilon^{T_{K}}=e^{T},\epsilon^{T_{I}}=e^{T_{i}%
},\epsilon^{T_{J}}=e^{T_{j}},$%

\begin{equation}
-\left[
\begin{array}
[c]{c}%
0\times\frac{1}{M}\left(  \frac{-t}{2}\right)  k_{4}^{T_{j}}-0\times\frac
{1}{M}\left(  \frac{-u}{2}\right)  k_{3}^{T_{j}}\\
0\times\frac{1}{M}\left(  \frac{-u}{2}\right)  k_{3}^{T_{i}}-0\times\frac
{1}{M}\left(  \frac{-t}{2}\right)  k_{4}^{T_{i}}%
\end{array}
\right]  =0.
\end{equation}

\textbf{Case three}: for $\epsilon^{T_{K}}=e^{T_{k}},\epsilon^{T_{I}}%
=e^{T},\epsilon^{T_{J}}=e^{T_{j}},$%
\begin{align}
&  -\left[
\begin{array}
[c]{c}%
0\times\frac{1}{M}\left(  \frac{-t}{2}\right)  k_{4}^{T_{j}}-0\times\frac
{1}{M}\left(  \frac{-u}{2}\right)  k_{3}^{T_{j}}\\
+\delta_{jk}\frac{1}{M}\left(  \frac{-u}{2}\right)  k_{3}^{T}-\delta_{jk}%
\frac{1}{M}\left(  \frac{-t}{2}\right)  k_{4}^{T}%
\end{array}
\right] \nonumber\\
&  =\delta_{jk}\frac{1}{2M}\left(  t+u\right)  k_{3}^{T}=\frac{-2E^{2}%
\delta_{jk}k_{3}^{T}}{M}.
\end{align}
In summary, we get the following ratios in the NS sector\begin{widetext}
\begin{align}
&  \left\langle b_{\frac{-1}{2}}^{T},\alpha_{-1}^{T}b_{\frac{-1}{2}}%
^{T}\right\rangle :\left\langle b_{\frac{-1}{2}}^{T},\alpha_{-1}^{L}%
b_{\frac{-1}{2}}^{L}\right\rangle :\left\langle b_{\frac{-1}{2}}^{T}%
,b_{\frac{-3}{2}}^{L}\right\rangle :\left\langle b_{\frac{-1}{2}}^{T_{i}%
},\alpha_{-1}^{T}b_{\frac{-1}{2}}^{T_{j}}\right\rangle :\left\langle
b_{\frac{-1}{2}}^{T_{k}},b_{\frac{-1}{2}}^{L}b_{\frac{-1}{2}}^{T}b_{\frac
{-1}{2}}^{T_{l}}\right\rangle \nonumber\\
= &  -2k_{3}^{T}E^{2}B:\frac{-2k_{3}^{T}E^{2}B}{M^{2}}:\frac{2k_{3}^{T}E^{2}%
B}{M}:\delta_{ij}2k_{3}^{T}E^{2}B:\delta_{lk}\frac{-2k_{3}^{T}E^{2}}%
{M}B=1:\frac{1}{M^{2}}:-\frac{1}{M}:-\delta_{ij}:\frac{\delta_{lk}}%
{M}.\label{omit}%
\end{align}
\end{widetext}where we have, for simplicity, omitted the last two tachyon
vertices in the notation of each HSSA in Eq.(\ref{omit}), which are the
superstring version of the bosonic string ratios in Eq.(\ref{bosonic}).
However, there is no inter-particle gauge transformation for the NS sector of
the superstring case as in the analysis of section II.

At this stage, one might think that there are $1+1+1+7+7=17$ HSSA in
Eq.(\ref{omit}) which is inconsistent with the $2^{4}=16$ hard PFSSA
calculated in \cite{PFSSA}. To resolve the inconsistency, one notes that
$b_{\frac{-3}{2}}^{L}$ in the third amplitude in Eq.(\ref{omit}) cannot be
alone a (high energy) physical state. So one needs to do a linear combination
of the first three amplitudes in Eq.(\ref{omit}) to transform the HSSA from
the operator basis to the physical state basis. The first physical state
amplitude can be obtained by choosing
\begin{equation}
\text{\ }\epsilon_{\left(  \mu\nu\right)  }=e_{\mu}^{T}e_{\nu}^{T}%
,\epsilon_{\nu}=-\epsilon_{\left(  \mu\nu\right)  }k^{\mu}=0
\end{equation}
in Eq.(\ref{6}) to obtain%
\begin{equation}
\left(  \epsilon_{(\mu\nu)}b_{\frac{-1}{2}}^{\mu}\alpha_{-1}^{\nu}%
+\epsilon_{\mu}b_{\frac{-3}{2}}^{\mu}\right)  \left\vert 0,k\right\rangle
\rightarrow\left(  b_{\frac{-1}{2}}^{T}\alpha_{-1}^{T}\right)  \left\vert
0;k\right\rangle \rightarrow1.
\end{equation}
The second physical state amplitude can be obtained by choosing%

\begin{align}
\epsilon_{\nu}  &  =-\epsilon_{\left(  \mu\nu\right)  }k^{\mu}=\frac{1}%
{2}e_{\mu}^{P}e_{\nu}^{L}k^{\mu}=\frac{-M}{2}e_{\nu}^{L},\nonumber\\
\epsilon_{\left(  \mu\nu\right)  }  &  =\frac{1}{2}\left(  e_{\mu}^{P}e_{\nu
}^{L}+e_{\nu}^{P}e_{\mu}^{L}\right)
\end{align}
in Eq.(\ref{6}) to obtain%
\begin{align}
&  \left(  \epsilon_{\left(  \mu\nu\right)  }b_{\frac{-1}{2}}^{\mu}\alpha
_{-1}^{\nu}+\epsilon_{\mu}b_{\frac{-3}{2}}^{\mu}\right)  \left\vert
0;k\right\rangle \nonumber\\
\rightarrow &  \left(  b_{\frac{-1}{2}}^{P}\alpha_{-1}^{L}+b_{\frac{-1}{2}%
}^{L}\alpha_{-1}^{P}-Mb_{\frac{-3}{2}}^{L}\right)  \left\vert 0;k\right\rangle
\nonumber\\
=  &  \left(  2b_{\frac{-1}{2}}^{L}\alpha_{-1}^{L}-Mb_{\frac{-3}{2}}%
^{L}\right)  \left\vert 0;k\right\rangle \rightarrow\frac{2}{M^{2}}+1.
\end{align}
So the ratios of the hard SSA in the NS sector at this mass level
are\begin{widetext}
\begin{align}
&  \left\langle b_{\frac{-1}{2}}^{T},\alpha_{-1}^{T}b_{\frac{-1}{2}}%
^{T}\right\rangle :\left\langle b_{\frac{-1}{2}}^{T},\left(  2b_{\frac{-1}{2}%
}^{L}\alpha_{-1}^{L}-Mb_{\frac{-3}{2}}^{L}\right)  \right\rangle :\left\langle
b_{\frac{-1}{2}}^{T_{i}},\alpha_{-1}^{T}b_{\frac{-1}{2}}^{T_{j}}\right\rangle
:\left\langle b_{\frac{-1}{2}}^{T_{k}},b_{\frac{-1}{2}}^{L}b_{\frac{-1}{2}%
}^{T}b_{\frac{-1}{2}}^{T_{l}}\right\rangle \nonumber\\
&  =-2k_{3}^{T}E^{2}B:-2(\frac{2}{M^{2}}+1)k_{3}^{T}E^{2}B:\delta_{ij}%
2k_{3}^{T}E^{2}B:\delta_{lk}\frac{-2k_{3}^{T}E^{2}}{M}B=1:2:-\delta_{ij}%
:\frac{\delta_{lk}}{\sqrt{2}}.\label{susy}
\end{align}
\end{widetext}In sum, in the NS sector one gets%
\begin{equation}
1+1+7+7=16
\end{equation}
HSSA in Eq.(\ref{susy}). This result agrees with those of hard massive PFSSA
in the R sector calculated recently \cite{PFSSA}.

\section{Discussion}

One main result of this work is the derivation of the superstring version of
Eq.(\ref{bosonic}), namely, the ratios in Eq.(\ref{omit}). The important
motivation to do this derivation is of course due to the recent calculation of
the hard PFSSA in \cite{PFSSA}.

Both the effects of "worldsheet fermion exchange" \cite{susy} and
$\mathit{no}$ "inter-particle gauge transformation" of the NS sector shown in
this paper are related to the existence of R sector for the case of
superstring. This means that it is important to include R sector \cite{PFSSA}
when considering high energy symmetry of superstring theory. Unfortunately,
the construction of the general massive fermion string vertex operators is
difficult at this stage. However, the results of this paper seem to indicate
that spacetime SUSY of the spectrum may help to provide useful information.%

\setcounter{equation}{0}
\renewcommand{\theequation}{\arabic{section}.\arabic{equation}}%

\begin{acknowledgments}
This work is supported in part by the Ministry of Science and Technology
(MoST) and S.T. Yau center of National Chiao Tung University (NCTU), Taiwan.
\end{acknowledgments}

\end{document}